\documentclass[aps,prd,twocolumn,showpacs,superscriptaddress,amsmath,graphicx]{revtex4}

\usepackage{graphicx}

\def\slashed#1{\protect{\slash\hspace{-5pt}#1}}

\begin{document}

\title{Charm meson resonances and $D \to V$ semileptonic form factors}

\author{Svjetlana Fajfer}
\email[Electronic address:]{svjetlana.fajfer@ijs.si}
\affiliation{Physik Department, Technische Universit\"at M\"unchen,
D-85748 Garching, Germany}
\affiliation{J. Stefan Institute, Jamova 39, P. O. Box 3000, 1001 Ljubljana, Slovenia}
\affiliation{Department of Physics, University of Ljubljana, Jadranska 19, 1000 Ljubljana, Slovenia}

\author{Jernej Kamenik}
\email[Electronic address:]{jernej.kamenik@ijs.si}
\affiliation{J. Stefan Institute, Jamova 39, P. O. Box 3000, 1001 Ljubljana, Slovenia}

\date{\today}

\begin{abstract}
Using limits of large energy effective theory and heavy quark effective theory we propose a simple parametrization of the heavy to light $H \to V$ semileptonic form factors. Then we reconsider $D \to V \ell \nu_{\ell}$ decays within a model which combines heavy meson and chiral symmetry. In our Lagrangians we include contributions coming from excited charm meson states, some of them recently observed. Within this framework we determine all parameters describing the shapes of the form factors and calculate branching ratios and helicity ratios for all $D \to V \ell \nu_{\ell}$ decays.
\end{abstract}


\maketitle

\section{Introduction}

Presently, one of the most important issues in hadronic physics is the extraction of the 
CKM parameters from exclusive decays. An essential ingredient in 
this approach is the knowledge of the form factors' shapes in heavy to 
light weak transitions. Usually, the attention has been devoted to $B$ decays and the determination of the phase of the $V_{ub}$ CKM matrix element. At the same time in the charm sector, the most accurate determination of the size of $V_{cs}$ and $V_{cd}$ matrix elements is not from a direct measurement, mainly due to theoretical uncertainties in the calculations of the relevant form factors' shapes.  
\par
Recently CLEO and FOCUS have published interesting results on 
$D^0 \to \pi^- \ell^+ \nu_{\ell}$ and  $D^0 \to K^- \ell^+ \nu_{\ell}$ decays~\cite{Link:2004dh,Huang:2004fr}. Their studies indicate that the single pole parametrization  of the relevant form factor cannot explain their data very well, leading to unphysical pole masses.
Both experimental groups have also attempted a modified pole fit, which was first put forward for $B$ decays in~\cite{Becirevic:1999kt}, and their results suggest the existence of 
contributions beyond lowest lying charm meson resonances. 
\par
On the other hand we have recently~\cite{Fajfer:2004mv} reconsidered $D \to P \ell \nu_{\ell}$ decay 
form factors within a framework which combines heavy  meson and chiral symmetries (HM$\chi$T) and includes in the interacting Lagrangian contributions coming from excited charm meson states. We have found that a two-poles shape of the relevant form factor can be successfully accommodated within HM$\chi$T when excited meson states are included into the model. In our approach the first pole is described by the lowest lying vector resonance, as in the original idea~\cite{Becirevic:1999kt}, while for the second one we assume complete saturation by the next vector state which we include in our Lagrangian. In doing this we anticipate the discrepancies from the general two-poles procedure, in which the second effective pole should account for all other excitations that might be exchanged in the $t$-channel, to be small and encoded in the parameters of the model. The unknown parameters have been obtained by fitting the experimental results for the branching ratio. The assumed pole behavior agrees well with experimental results confirming our anticipation of small saturation error. 
\par
In addition to studies of heavy to light pseudoscalar meson weak transitions ($H\to P$), transitions of heavy pseudoscalar mesons to light vector mesons ($H\to V$) such as $D_s\to \phi \ell \nu_{\ell}$ an $D_s\to K^* \ell \nu_{\ell}$ offer an opportunity to extract the size of the relevant CKM matrix elements.
We continue with our study and re-investigate vector and axial-vector form factors in $D \to V \ell \nu_{\ell}$ decays within a similar framework as in the case of $D\to P \ell \nu_{\ell}$. The $H\to V$ transitions were already carefully investigated within many different frameworks such as perturbative QCD~\cite{Kurimoto:2001zj,Mahajan:2004dx}, QCD sum rules~\cite{Ball:1993tp, Ball:2004rg, Bakulev:2000fb, Wang:2001bh, Du:2003ja, Aliev:2004qd}, lattice QCD~\cite{Flynn:1995dc, DelDebbio:1997kr, Demchuk:1997uz, Abada:2002ie}, a few attempts to use combined heavy meson and chiral Lagrangians (HM$\chi$T)~\cite{Bajc:1995km, Casalbuoni:1996pg}, quark models~\cite{Wirbel:1985ji, Scora:1995ty, Faustov:1995bf, Melikhov:2000yu}, large energy effective theory (LEET)~\cite{Charles:1998dr} and soft collinear effective theory (SCET)~\cite{Beneke:2000wa,Bauer:2000yr, Burdman:2000ku, Ebert:2001pc, Hill:2004if, Hill:2004rx}. Each of these approaches has only a limited range of validity. For example, the QCD sum rules, LEET and SCET are suitable only for the low $q^2$ region while lattice QCD and HM$\chi$T are successful for maximal $q^2$. It is important to note, that currently lattice QCD and QCD sum rules are the only approaches that enable the computation of form factors solely from first principles. On the other hand quark models usually involve parameters which have little physical correspondence to the underlying theory of QCD.
\par
The experimental situation in $D \to V \ell \nu_{\ell}$ has not changed a lot in the last few years, but recently it has been gaining pace~\cite{Link:2002wg,Link:2004qt,Link:2004gp,Link:2005ge,Coan:2005iu,Huang:2005iv}, and hopefully
more results on the $q^2$ shape of the form factors will be available soon. Unlike in the case of $H \to P$ weak transitions, no general parametrization of the form factors, relevant to $H \to V$ weak decays has yet been proposed. Usually a simple pole behavior of all the form factors is assumed when extracting values of the form factors at $q^2=0$ from experiment or extrapolating results of different theoretical approaches. 
\par
Recently, the spectroscopy of charm mesons has been enriched by discoveries of many  new charm meson resonances. BaBar~\cite{Aubert:2003fg} collaboration has announced
a new, narrow meson $D_{sJ}(2317)^+$. This was confirmed by
Focus~\cite{Vaandering:2004ix} and CLEO~\cite{Besson:2003jp} which
also noticed another narrow state, $D_{sJ}(2463)^+$. Both states were
confirmed by Belle~\cite{Krokovny:2003zq}. 
Finally, Selex~\cite{Evdokimov:2004iy} has announced a new, surprisingly narrow
state $D_{sJ}^ +(2632)$. The states $D_{sJ}(2317)^+$ and $D_{sJ}(2463)^+$ 
%have already been proposed as members of the $(0^+,1^+)$ spin multiplet chiral partners of the heavy-light  pseudoscalar and vector $D_s$ mesons~\cite{Bardeen:2003kt,Nowak:2004uv}
are already being identified to belong to the $(0^+,1^+)$ spin-parity doublet of the $D_s$ mesons while the $D_{sJ}^+(2632)$ state has been proposed as the first radial
excitation of the $D_s^*(2112)$ with the spin parity assignment $1^-$~\cite{Barnes:2004ay,vanBeveren:2004ve,Dai:2004ng}.
\par
The purpose of this study  is to (1) devise a general parametrization of all the form factors relevant to $H\to V$ weak transitions which would take into account known experimental results on heavy meson resonances as well as known theoretical limits of heavy quark effective theory (HQET) and LEET relevant to $H\to V$ weak  transitions; to (2) investigate contributions of the newly discovered charm mesons to $D\to V$ semileptonic decays within an effective model based on HM$\chi$T by incorporating the newly discovered heavy meson fields into the HM$\chi$T Lagrangian and utilizing the general form factor parametrization. We restrain our
discussion to the leading chiral and $1/m_H$ terms in the expansion,
but we hope to capture the main physical features about the impact of
the nearest poles in the $t$-channel to the $q^2$-dependence of the
form factors. 
\par
In Sec.~II we revise the common weak current matrix element decomposition relevant to transitions between pseudoscalar and vector mesons and introduce a form factor decomposition, which is independent of the mass of the pseudoscalar meson and thus convenient for studying $H\to V$ weak transitions. In Sec.~III we derive a general $H\to V$ form factor parametrization drawing from both known experimental properties of heavy mesons as well as from known theoretical scaling laws and form factor relations in the limit of the infinite heavy meson mass. Sec.~IV describes the framework we use in our HM$\chi$T calculations: we
write down the HM$\chi$T Lagrangian for for heavy and light mesons and
extend it to incorporate new heavy meson fields. In Sec.~V we calculate the values of the $D \to V$ semileptonic form factors near zero recoil within HM$\chi$T and extrapolate our results to larger recoils using the general parametrization of Sec.~III and by saturating the effective poles with physical masses of experimentally known or theoretically predicted charmed resonances. Finally, a short summary of the results and comparison with other approaches as well as with existing experimental data is given in Sec.~VI.

\section{Parametrization of $H \to V$ current matrix element}
 
A frequently encountered decomposition of the current matrix elements relevant to semileptonic decays between a heavy pseudoscalar meson state $|H(p_H)\rangle$ with momentum $p_H^{\nu}$ and a light vector meson state $|V(p_V,\epsilon_V)\rangle$ with momentum $p_V^{\nu}$ and polarization vector $\epsilon_V^{\nu}$ is
\begin{widetext}
\begin{eqnarray}
	\langle V (\epsilon_V,p_V) | \bar q \gamma^{\mu} Q | H (p_H) \rangle &=& \frac{2 V(q^2)}{m_H + m_V} \epsilon^{\mu\nu\alpha\beta} \epsilon_{V\nu}^* p_{H\alpha} p_{V\beta}, \nonumber\\*
	\langle V (\epsilon_V,p_V) | \bar q \gamma^{\mu} \gamma^5 Q | H (p_H) \rangle &=&  - i \epsilon_V^* \cdot q \frac{2m_V}{q^2} q^{\mu} A_0(q^2) - i(m_H + m_V) \left[\epsilon_V^{*\mu} - \frac{\epsilon_V^{*} \cdot q}{q^2} q^{\mu}\right] A_1(q^2) \nonumber\\*
	&&+ i \frac{\epsilon_V^{*}\cdot q}{(m_H + m_V)}\left[(p_H+p_V)^{\mu} - \frac{m_H^2-m_V^2}{q^2} q^{\mu}\right] A_2(q^2),\nonumber\\* 
\end{eqnarray}
\end{widetext}
where $q=u,d$~or~$s$ are the light quark fields, $Q=b$~or~$c$ denote the heavy quark fields, $q^{\nu} = (p_H-p_V)^{\nu}$ is the exchanged momentum and $q^2$ is the exchanged momentum squared. Here $V$ denotes the vector form factor and is expected to be dominated by vector meson resonance exchange, the axial $A_1$ and $A_2$ form factors are expected to be dominated by axial resonances, while $A_0$ denotes the pseudoscalar form factor and is expected to be dominated by pseudoscalar meson resonance exchange~\cite{Wirbel:1985ji}.
In order that these matrix elements are finite at $q^2=0$, the form factors must also satisfy the well known relation
\begin{equation}
A_0(0)-\frac{m_H+m_V}{2m_V}A_1(0)+\frac{m_H-m_V}{2m_V}A_2(0)=0~.
\label{PV_form_factor_relations}
\end{equation}
\par
We will work in the static limit of HQET where the eigenstates of QCD and HQET Lagrangians are related as
\begin{equation}
\lim_{m_H\to\infty} \frac{1}{\sqrt m_H} |H(p_H)\rangle_{QCD} = | H(v)\rangle_{HQET}.
\end{equation}
In this limit it is more convenient to use definitions in which the form factors are independent of the heavy meson mass, namely we propose
\begin{eqnarray}
	\langle V (\epsilon_V,p_V) | \bar q \gamma^{\mu} Q_v | H (v) \rangle &=& f_v \epsilon^{\mu\nu\alpha\beta} \epsilon_{V\nu}^* v_{\alpha} p_{V\beta}, \nonumber\\* 
	\langle V (\epsilon_V,p_V) | \bar q \gamma^{\mu}\gamma^{5} Q_v | H (v) \rangle &=& - i a_2 (\epsilon^*_V \cdot v) \left[p^{\mu}_V - (v\cdot p_V) v^{\mu}\right]\nonumber\\*
	&& - i a_1 \left[ \epsilon^{*\mu}_V - (v \cdot \epsilon_V^*) v^{\mu} \right]\nonumber\\*
	&& - i a_0 (v\cdot \epsilon^{*}_V) v^{\mu},
\end{eqnarray}
where the HQET heavy quark field $Q_v$ is independent of the heavy quark mass. The form factors $f_v$, $a_1$, $a_2$ and $a_0$ are functions of the variable
\begin{equation}
v \cdot p_V = \frac{m_H^2 + m_V^2 - q^2}{2m_H},
\end{equation}
which in the heavy meson rest frame is the energy of the light meson $E_V$. In such decomposition, all the form factors ($f_v$, $a_1$, $a_2$ and $a_0$) scale as constants with the heavy meson mass. The relation between the two form factor decompositions is obtained by correctly matching QCD and HQET at the scale $\mu \sim m_Q$~\cite{Eichten:1989zv,Broadhurst:1994se}: 
\begin{widetext}
\begin{eqnarray}
	\frac{C_{\gamma_1}(m_Q)}{\sqrt m_H} \left[ f_V(v\cdot p_V) + \mathcal O(1/m_H) \right] &=&  \frac{2 V(q^2)}{m_H+m_V}|_{q^2\approx q^2_{\mathrm{max}}},\nonumber\\*
	\frac{C_{\gamma_0\gamma_5}(m_Q)}{ \sqrt{m_H}} \left[ a_0(v\cdot p_V) +\mathcal O(1/m_H)\right] &=&  \Bigg\{ \frac{(m_H-E_V)}{q^2} \left[ 2 m_V A_0(q^2) + (m_H+m_V) A_1(q^2) - (m_H-m_V) A_2(q^2) \right]\nonumber\\*
	&& + \frac{(m_H+E_V)}{m_H+m_V} A_2(q^2) - \frac{(m_H+m_V)}{m_H} A_1(q^2) \Bigg\} |_{q^2\approx q^2_{\mathrm{max}}},\nonumber\\*
	C_{\gamma_1\gamma_5}(m_Q) \sqrt{m_H} \left[a_1(v\cdot p_V) + \mathcal O(1/m_H) \right] &=&  (m_H+m_V)  A_1(q^2) |_{q^2\approx q^2_{\mathrm{max}}},\nonumber\\*
	\frac{C_{\gamma_1\gamma_5}(m_Q)}{\sqrt m_H} \left[a_2(v\cdot p_V) + \mathcal O(1/m_H) \right] &=& \Big\{ \frac{m_H+m_V}{q^2} \left[A_1(q^2) + A_0(q^2)\right] - \frac{m_H-m_V}{q^2} \left[ A_2(q^2) + A_0(q^2)\right] \nonumber\\*
	&& - \frac{A_2(q^2)}{m_H+m_V} \Big\} |_{q^2\approx q^2_{\mathrm{max}}}.
\end{eqnarray} 
\end{widetext}
\par
In the following we set the matching constants $C_{\Gamma}$ to their tree level values $(C_{\Gamma}=1)$. At leading order in $1/m_Q$ we thus get
\begin{eqnarray}
V(q^2)|_{q^2\approx q^2_{\mathrm{max}}} &=& \frac{\sqrt m_H}{2 } f_v (v \cdot p_V), \nonumber\\* 
A_1(q^2)|_{q^2\approx q^2_{\mathrm{max}}} &=& \frac{1}{\sqrt m_H} a_1 (v \cdot p_V), \nonumber\\*
A_2(q^2)|_{q^2\approx q^2_{\mathrm{max}}} &=& \frac{\sqrt m_H}{2} a_2 (v \cdot p_V), \nonumber\\*
A_0(q^2)|_{q^2\approx q^2_{\mathrm{max}}} &=& \frac{\sqrt m_H}{2 m_V } a_0 (v \cdot p_V), 
\label{eq_HQET_scaling}
\end{eqnarray}
which exhibit the usual heavy meson mass scaling laws for the semileptonic form factors~\cite{Isgur:1990kf}. This parametrization is especially useful when calculating the form factors within HM$\chi$T. The individual contributions of different terms in the HM$\chi$T Lagrangian to various form factors can be easily projected out.

\section{Parametrization of the form factors}

Next we propose a general parametrization of the heavy to light vector form factors, which takes into account all the known scaling and resonance properties of the form factors.
As already evident from Eq.~(\ref{eq_HQET_scaling}), there exist the well known HQET scaling laws in the limit of zero recoil~\cite{Isgur:1990kf}. On the other hand in the large energy limit $q^2\to 0$, one obtains the following expressions for the form factors~\cite{Charles:1998dr}
\begin{eqnarray}
V(q^2)|_{q^2\approx 0} &=& \frac{m_H + m_V}{m_H} \xi_{\perp} (E_V), \nonumber\\* 
A_1(q^2)|_{q^2\approx 0} &=& \frac{2 E_V} {{m_H + m_V}} \xi_{\perp} (E_V), \nonumber\\* 
A_2(q^2)|_{q^2\approx 0} &=& \frac{m_H + m_V}{m_H} \left[ \xi_{\perp} (E_V) - \frac{m}{E} \xi_{\parallel} (E_V) \right],\nonumber\\*
A_0(q^2)|_{q^2\approx 0} &=& \left( 1 - \frac{m_V^2}{2 E_V m_H} \right) \xi_{\parallel} (E_V) \approx \xi_{\parallel} (E_V),
\label{eq_LEET_scaling}
\end{eqnarray}
where both universal LEET functions $\xi_{\perp}$ and $\xi_{\parallel}$ scale with the heavy meson mass as $m_H^{-3/2}$. These scaling laws were subsequently confirmed by means of SCET~\cite{Beneke:2000wa, Ebert:2001pc}. This is important since the LEET description breaks down beyond the tree level due to missing soft gluonic degrees of freedom which are however systematically taken into account within SCET.
\par
The starting point is the vector form factor $V$, which is dominated by the pole at $t=m_{H^*}^2$ when considering the part of the phase space that is close to the zero recoil. %This physical pole is known not only experimentally~\cite{?} but was also obtained in lattice QCD simulations~\cite{?}. 
It is very easy to see, that the residuum at that pole
%\begin{equation}
%\mathrm{Res}_{q^2=m_{H^*}^2}V(q^2) = \frac{1}{2}g_{H^* H V} f_{H^*} (m_H+m_V)
%\end{equation}
scales as $\sim m_H^{3/2}$ with the heavy meson mass~\cite{Becirevic:1999kt}. For the heavy to light transitions this situation is expected to be realized near the zero recoil where also the HQET scaling~(\ref{eq_HQET_scaling}) applies. However, since the kinematically accessible region $q^2\in(0,q^2_{\mathrm{max}}]$ is large, the pole dominance can be used only on a small fraction of the phase space, $i.e.$ for $|\vec q|\approx 0$. Even in this region the situation for $H\to V$ form factors is more complex than in the case of $H\to P$ transitions, where $q^2_{\mathrm{max}}$ is indeed very close to the vector pole due to low mass of the light pseudoscalar mesons. Here, due to larger masses of the light vector mesons, $q^2_{\mathrm{max}}$ is pushed away from the resonance pole and the form factor may not be completely saturated by it. 
%Consequently in these transitions even the first pole may appear shifted in relation to the physical pole mass of the first heavy vector resonance meson. 
For the sake of clarity and conciseness we, however, in our present study neglect such possible discrepancies and assume complete saturation of the vector form factor in this region by the first physical resonance. On the other hand, in the region of large recoils, LEET dictates the scaling~(\ref{eq_LEET_scaling}). In the full analogy with the discussion made in Refs. \cite{Becirevic:1999kt, Hill:2005ju}, the vector form factor consequently receives contributions from two poles and can be written as
\begin{equation}
V(q^2) = c'_H \frac{1-a}{(1-x)(1-a x)},
\label{eq_v_ff}
\end{equation}
where $x=q^2/m_{H^*}^2$ ensures, that the form factor is dominated by the physical $H^*$ pole, while $a$ measures the contribution of higher states which are parametrized by another effective pole at $m_{\mathrm{eff}}^2=m_{H^*}^2/a$. The parameters $c'_H$ and $a$ scale with the heavy meson mass as $c'_H\sim m_H^{-1/2}$ and $a\sim 1 - a_0/m_{H}$ to ensure the correct form factor scaling in both small and large recoil regions. 
\par
An interesting and useful feature one gets from the large energy limit is the relation between $V$ and $A_1$~\cite{Charles:1998dr}
\begin{equation}
\left[ V(q^2)/A_1(q^2) \right]|_{q^2\approx 0} = \frac{(m_H+m_V)^2}{2 E_V m_H},
\label{eq_VA_LEET}
\end{equation}
which is valid up to terms $\propto 1/m_H^2$~\cite{Ebert:2001pc}. This relation remains valid even when the leading order corrections due to soft gluon exchange are taken into account~\cite{Burdman:2000ku, Hill:2004rx}. When combined with our result~(\ref{eq_v_ff}), it imposes a single pole structure on $A_1$. We can thus continue in the same line of argument and write
\begin{equation}
A_1(q^2) = c'_H \xi \frac{1-a}{1-b' x}.
\label{eq_a1_ff}
\end{equation}
Here $\xi=m_H^2/(m_H+m_V)^2$ is the proportionality factor between $A_1$ and $V$ from~(\ref{eq_VA_LEET}), while $b'$ measures the contribution of higher states with spin-parity assignment $1^+$ which are parametrized by the effective pole at $m_{H'^*_{\mathrm{eff}}}^2=m_{H^*}^2/b'$. It can be readily checked that also $A_1$, when parametrized in this way, satisfies all the scaling constraints. 
\par
Next we parametrize the $A_0$ form factor, which is completely independent of all the others so far as it is dominated by the pseudoscalar pole and is proportional to a different universal function in LEET. To satisfy both HQET and LEET scaling laws we parametrize it as 
\begin{equation}
A_0(q^2) = c''_H \frac{1-a'}{(1-y)(1-a' y)},
\label{eq_a0_ff}
\end{equation}
where $y = q^2/m_H^2$ ensures the physical $0^-$ pole dominance at small recoils. Imposing $c''_H\sim m_H^{-1/2}$ and $a'\sim 1-a'_0/m_H$ preserves all scaling laws, while $a'$ again parametrizes the contribution of higher pseudoscalar states by an effective pole at $m_{H'_{\mathrm{eff}}}^2=m_{H}^2/a'$. The resemblance to $V$ is obvious and due to the same kind of analysis~\cite{Becirevic:1999kt} although the parameters appearing in the two form factors are completely unrelated. 
\par
Finally for the $A_2$ form factor, due to the pole behavior of the $A_1$ form factor on one hand and different HQET scaling at $q^2_{\mathrm{max}}$~(\ref{eq_HQET_scaling}) on the other hand, we have to go beyond a simple pole formulation. Thus we impose
\begin{equation}
A_2(q^2) = \frac{(m_H+m_V) \xi c'_H (1-a) + 2 m_V c''_H (1-a')}{(m_H-m_V)(1-b' x)(1-b'' x)},
\label{eq_a2_ff}
\end{equation}
which again satisfies all constraints. Due to the relation~(\ref{PV_form_factor_relations}) we only gain one new parameter in this formulation, $b''$. This however causes the contribution of the $1^+$ resonances to be shared between the two effective poles in this form factor.
\par
At the end we have parametrized the four $H\to V$ vector form factors in terms of the six parameters $c'_H$, $a$, $b'$, $a'$, $c''_H$ and $b''$.
\par
It is convenient to introduce helicity amplitudes for the decays $H\rightarrow V \ell \nu$ as in for example~\cite{Ball:1991bs}:
\begin{eqnarray}
H_{\pm}(y) &=& + (m_H+m_V) A_1(m_H^2 y) \mp \frac{2 m_H |\vec p_V(y)|}{m_H+m_V} V(m_H^2 y)\nonumber\\*
H_0(y) &=& + \frac{m_H+m_V}{2 m_H m_V \sqrt y} [ m_H^2 (1-y) -m_V^2] A_1(m_H^2 y)\nonumber\\*
	&& - \frac{2 m_H |\vec p_V(y)|}{m_V(m_H+m_V) \sqrt y } A_2(m_H^2 y)
\end{eqnarray} 
where $y=q^2/m_H^2$ and the three-momentum of the light vector meson is given by: 
\begin{equation}
|\vec p_V (y)|^2 = \frac{[m_H^2 (1-y) + m_V^2]^2}{4 m_H^2} -m_V^2.
\end{equation}
As shown in Ref.~\cite{Ebert:2001pc} these helicity amplitudes can be related to individual form factors near $q^2=0$. Using relations~(\ref{eq_LEET_scaling}), valid in the large energy limit, one can write
\begin{eqnarray}
H_{-}(y)|_{y\approx 0} &\approx& 2(m_H+m_V) A_1(m_H^2 y),\nonumber\\  
H_{+}(y)|_{y\approx 0} &\simeq& 0,  
\end{eqnarray}
but also, by using relation~(\ref{PV_form_factor_relations})
\begin{equation}
H_0(y)|_{y\approx 0} \approx \frac{2 |\vec p_V(y)|}{\sqrt{y}} A_0(m_H^2 y).
\label{eq_H0_A0}
\end{equation}
Thus in this region we can probe directly for the parameters $c'_H(1-a)$ and $c''_H(1-a')$.
\par
On the other hand in the region of small recoil ($|\vec p_V| \simeq 0$ or $y\approx y_{\mathrm{max}}$) the helicity amplitudes are saturated by the $A_1$ form factor
\begin{eqnarray}
H_{\pm}(y)|_{y\approx y_{\mathrm{max}}} &\approx& (m_H+m_V) A_1(m_H^2 y), \nonumber\\
H_{0}(y)|_{y\approx y_{\mathrm{max}}} &\approx& -2 (m_H+m_V) \frac{m_V}{m_H} A_1(m_H^2 y).
\end{eqnarray}
Consequently we can also directly probe for the value of the $b'$ parameter determining the position of the first effective axial resonance pole by taking a ratio of $H_-$ helicity amplitude values at small and large recoils
\begin{equation}
\frac{H_{-}(y)|_{y\approx 0}}{H_{-}(y)|_{y\approx y_{\mathrm{max}}}} \approx 2 \left[ 1 - b' (m_H-m_V)^2/m_H^2 \right].
\end{equation}

\section{The Model}

\subsection{Strong interactions}

At leading order in chiral and $1/m_H$ expansion, strong interactions between lowest lying pseudoscalar and vector heavy meson fields, and light vector meson fields are described by the interaction Lagrangian~\cite{Bajc:1995km,Casalbuoni:1996pg}
\begin{equation}
\mathcal L_{\mathrm{int}} = - i \beta \langle H_b v_{\mu} \hat\rho^{\mu}_{ba} \bar H_a \rangle + i \lambda \langle H_b \sigma^{\mu\nu} F_{\mu\nu}(\hat\rho)_{ba} \bar H_a \rangle,
\end{equation} 
where the first term is even under parity transformation, and the second term is parity odd. $H=1/2(1+\slashed v )[P^*_{\mu} \gamma^{\mu} - P \gamma_5]$ is the matrix representation of the heavy meson fields, where $P^*_{\mu}$ and $P$ are creation operators for heavy vector and pseudoscalar mesons respectively. Light vector meson fields are described by $\hat\rho_{\mu} = i \frac{g_V}{\sqrt{2}} \rho_{\mu}$, where $\rho_{\mu}$ is the light vector meson field matrix
\begin{equation}
\rho_{\mu} = 
   \begin{pmatrix}
    \frac{1}{\sqrt 2} (\omega_{\mu} + \rho^0_{\mu}) & \rho^+_{\mu} & K^{*+}_{\mu} \\
    \rho^-_{\mu} & \frac{1}{\sqrt 2} (\omega_{\mu} - \rho^0_{\mu}) & K^{*0}_{\mu} \\
    K^{*-}_{\mu} & \bar K^{*0}_{\mu} & \phi_{\mu}
   \end{pmatrix}.
\end{equation}
The gauge field tensor is defined as $F_{\mu\nu}(\hat\rho) = \partial_{\mu} \hat\rho_{\nu} - \partial_{\nu} \hat\rho_{\mu} + [\hat \rho_{\mu} , \hat \rho_{\nu}]$. Furthermore, $\langle\ldots\rangle$ indicate a trace over spinor matrices and summation over light quark flavor indexes. 
\par
In order to incorporate positive parity heavy meson states into the model, we introduce the scalar-axial field multiplet $G=1/2(1+\slashed v )[S^*_{\mu} \gamma^{\mu}\gamma_{5} - S]$ representing axial ($S^*_{\mu}$) and scalar ($S$) mesons and incorporate it into the interaction Lagrangian by adding additional leading order interaction terms between heavy even and odd parity fields and light vector fields: 
\begin{eqnarray}
\mathcal L'_{\mathrm{int}} &=&  - i \zeta \langle H_b v_{\mu} \hat\rho^{\mu}_{ba} \bar G_a \rangle + \mathrm{h.c.} \nonumber\\*
&& + i \mu \langle H_b \sigma^{\mu\nu} F_{\mu\nu}(\hat\rho)_{ba} \bar G_a \rangle + \mathrm{h.c.}.
\label{L_even_odd}
\end{eqnarray} 
There exists another field multiplet in HM$\chi$T containing positive parity heavy meson states, $T^{\mu} = 1/2 (1+\slashed v) [ T_1^{\mu\nu}\gamma_{\nu} - \sqrt{3/2} T_{2\nu} \gamma_5 (g^{\mu\nu} -1/3 \gamma^{\nu}(\gamma^{\mu}-v^{\mu})) ]$, where $T_1^{\mu\nu}$ is the tensor field with spin-parity assignment $2^+$, while $T_2^{\mu}$ is another $1^+$ axial vector meson field. However, as pointed out in Ref.~\cite{Casalbuoni:1996pg}, the matrix element of the HM$\chi$T bosonized currents containing these fields between a single heavy meson state $|H(p_H)\rangle$ and the vacuum vanishes at leading order in $1/m_H$ due to heavy quark spin symmetry. Consequently, such fields do not contribute at leading order to $H\to V$ semileptonic decays.
\par
Finally we also want to include the radially excited states into our discussion and therefore introduce another odd parity heavy meson multiplet field $H'=1/2(1+\slashed v )[P^{'*}_{\mu} \gamma^{\mu} - P' \gamma_5]$ containing the radial excitations of ground state pseudoscalar and vector mesons. Such excited states were predicted in~\cite{DiPierro:2001uu}. The strong interactions between these fields, ground state heavy meson fields $H$ and light vector fields can again be described by the lowest order interaction Lagrangian analogous to~(\ref{L_even_odd})
\begin{eqnarray}
\tilde{\mathcal L}_{\mathrm{int}} &=& - i \widetilde \zeta \langle \widetilde H_b v_{\mu} \hat\rho^{\mu}_{ba} \bar {H}_a \rangle \nonumber\\*
&& + i \widetilde \mu \langle \widetilde H_b \sigma^{\mu\nu} F_{\mu\nu}(\hat\rho)_{ba} \bar {H}_a \rangle +\mathrm{h.c.}.
\label{L_odd_oddstar}
\end{eqnarray} 

\subsection{Weak interactions}

For the semileptonic decays the weak Lagrangian can be given by the effective current-current Fermi interaction
\begin{equation}
\mathcal L_{\mathrm{eff}} = - \frac{G_F}{\sqrt 2} \left[ \bar \ell \gamma^{\mu} (1-\gamma^5) \nu_{\ell} \mathcal J_{\mu}  \right], 
\end{equation} 
where $G_F$ is the Fermi constant and $\mathcal J$ is the effective hadronic current. In heavy to light meson decays it can be written as $\mathcal J= K^a J_a$, where constants $K^a$ parametrize the $SU(3)$ flavor mixing, while the leading order weak current $J_a$ in chiral and $1/m_H$ expansion between heavy ground state pseudoscalar and vector mesons and light vector mesons can be written as~\cite{Bajc:1995km,Casalbuoni:1996pg}
\begin{eqnarray}
J_a^{\mu} &=& \frac{1}{2} i \alpha \langle \gamma^{\mu} (1-\gamma^5) H_a \rangle \nonumber\\*
&& + \alpha_1 \langle \gamma^5 H_b \hat\rho^{\mu}_{ba} \rangle + \alpha_2 \langle \gamma^{\mu} \gamma^{5} H_b v_{\alpha} \hat\rho^{\alpha}_{ba} \rangle.
\end{eqnarray}
For our leading order calculation, we will also need the weak current operator for the scalar and axial heavy mesons' transition to the vacuum    
\begin{equation}
J_a^{'\mu} = \frac{1}{2} i \alpha' \langle \gamma^{\mu} (1-\gamma^5) G_a \rangle,
\end{equation}
and the same for radially excited pseudoscalar and vector fields
\begin{equation}
\tilde J_a^{\mu} = \frac{1}{2} i \tilde\alpha \langle \gamma^{\mu} (1-\gamma^5) H'_a \rangle.
\end{equation}

\section{Form Factor Calculation}

\subsection{HM$\chi$T calculation at zero recoil}

In HM$\chi$PT the derived Feynman rules are valid near zero recoil ($|\vec p_V|\simeq0$). For the heavy meson propagators we use $i\delta_{ab}/2(v\cdot k-\Delta)$ and $-i \delta_{ab} (g_{\mu\nu}-v_{\mu} v_{\nu})/2(v\cdot k - \Delta)$ for the pseudoscalar(scalar) and vector (axial)mesons respectively, where $k^{\mu}=q^{\mu}-m_{H} v^{\mu}$. $\Delta = \Delta_R$ is the mass splitting between the heavy resonance meson $R$ and the ground state strangeless heavy pseudoscalar meson. It comes from leading order $1/m_H$ (spin symmetry breaking), chiral and $SU(3)$ breaking corrections, when all heavy meson fields are normalized to physical masses of ground state strangeless heavy pseudoscalar mesons. For the hadronic current matrix element we thus get
\begin{eqnarray}
&&\langle V(p_V) | J^{\mu} | H(v) \rangle = - i \sqrt 2 g_V \left( \alpha_1 \epsilon_{V}^{\mu} - \alpha_2 v \cdot \epsilon_{V} v^{\mu} \right) \nonumber\\*
&& -\sqrt 2 g_V \alpha  \frac{\lambda \epsilon^{\mu\nu\alpha\beta} v_{\nu} p_{V\alpha} \epsilon_{V\beta}}{v\cdot p_V + \Delta_{H^*}} -\sqrt 2 g_V \widetilde \alpha  \frac{\widetilde\mu \epsilon^{\mu\nu\alpha\beta} v_{\nu} p_{V\alpha} \epsilon_{V\beta}}{v\cdot p_V + \Delta_{H'^*}} \nonumber\\*
&& - i \frac{g_V}{\sqrt 2} \alpha \frac{\beta v \cdot \epsilon_V v^{\mu} }{v\cdot p_V + \Delta_{H_P}} - i \frac{g_V}{\sqrt 2} \widetilde\alpha  \frac{\widetilde\zeta v \cdot \epsilon_V v^{\mu}}{v\cdot p_V + \Delta_{H'_P}} \nonumber\\*
&& - i \frac{g_V}{\sqrt 2} \alpha' \frac{\epsilon_V^{\mu} \left( \zeta - 2\mu v\cdot p_V \right) + \left( 2\mu p_V^{\mu} - \zeta v^{\mu} \right) v\cdot \epsilon_V }{v\cdot p_V + \Delta_{H_A}}.
\label{J_HMCPT}
\end{eqnarray} 
From this we extract the form factors $V(q^2)$, $A_1(q^2)$, $A_2(q^2)$ and $A_0(q^2)$:
\begin{eqnarray}
V(q^2)|_{q^2\approx q^2_{\mathrm{max}}} &=& -\frac{g_V}{\sqrt 2} \alpha m_H \sqrt m_H \frac{\lambda }{v\cdot p_V + \Delta_{H^*}} \nonumber\\*
&& - \frac{g_V}{\sqrt 2} \widetilde\alpha m_H \sqrt m_H \frac{\widetilde \mu }{v\cdot p_V + \Delta_{H'^*}} \nonumber\\*
A_1(q^2)|_{q^2\approx q^2_{\mathrm{max}}} &=& \frac{g_V}{\sqrt 2} \alpha' \frac{\sqrt m_H}{m_H + m_V} \frac{ \zeta - 2\mu v\cdot p_V }{v\cdot p_V + \Delta_{H_A}} \nonumber\\*
&& - \sqrt 2 g_V \alpha_1 \frac{\sqrt m_H}{m_H +m_V} \nonumber\\*
A_2(q^2)|_{q^2\approx q^2_{\mathrm{max}}} &=& \frac{g_V}{\sqrt 2} \alpha' \frac{m_H +m_V}{\sqrt m_H} \frac{\mu}{v\cdot p_V + \Delta_{H_A}} \nonumber\\*
A_0(q^2)|_{q^2\approx q^2_{\mathrm{max}}} &=& \frac{g_V}{2\sqrt 2} \frac{\sqrt m_H}{m_V} \Big( 2 \alpha_1 - 2 \alpha_2 \nonumber\\*
&& + \alpha \frac{\beta}{v\cdot p_V + \Delta_{H_P}} + \widetilde \alpha \frac{\widetilde\zeta}{v\cdot p_V + \Delta_{H'_P}} \Big)\nonumber\\*
\label{eq_ff_HMcT}
\end{eqnarray}

\subsection{Extrapolation to higher recoils}

In order to extrapolate our HM$\chi$T calculation results at $q^2\approx q^2_{\mathrm{max}}$ to higher recoils we employ the general analysis from Sec.~III. We model the form factors' $q^2$ behavior using the formulas~(\ref{eq_v_ff}),~(\ref{eq_a1_ff}),~(\ref{eq_a0_ff}) and~(\ref{eq_a2_ff}) with model matching conditions at $q^2_{\mathrm{max}}$. In order to reduce the number of free parameters in this extrapolation we employ the same strategy as in our previous work~\cite{Fajfer:2004mv}. We use the information on the contributions of different resonances to the form factors as suggested by our model. For the vector form factor $V$ we thus propose $a=m_{H^*}^2/m_{H'^*}^2$ which saturates the effective second pole by the first vector radial excitation $H'^*$. Similarly we set $b'=m_{H^*}^2/m_{H_A}^2$ and $a'=m_{H}^2/m_{H'}^2$ saturating the poles of the $A_0$ and $A_1$ form factors and the first pole of the $A_2$ form factor with the $H'$ pseudoscalar radial excitation and the $H_A$ orbital axial excitation respectively. Since our model does not contain a second resonance contribution to the $A_2$ form factor, we impose $b''=0$, effectively sending the second pole mass of this form factor to infinity. 
\par
At the end we have fixed all the pole parameters appearing in the general form factor parametrization formulas of Sec.~III using physical information and model predictions on the resonances contributing to the various form factors. The remaining parameters ($c_H$ and $c'_H$) are on the other hand related to the parameters of HM$\chi$T via the model matching conditions at zero recoil.

\section{Numerical results}

We now leave the general discussion of $H\to V$ transitions and restrict our present study to $D$ decays, although our calculations can readily be applied to semileptonic decays of $B$ mesons once more experimental information becomes available on excited $B$ meson resonances. In our numerical analysis we use available experimental information and theoretical predictions on charm meson resonances. Particularly for the $D_s$
axial resonance we use the mass $m_{D_{sJ}(2460)}=2.459~\mathrm{GeV}$, while for the first orbital excitation of the $D^*$ meson, we use the mass of $m_{D_1(2420)}=2.422~\mathrm{GeV}$~\cite{Eidelman:2004wy}. For the radially excited vector resonance we then have the Selex $D_{sJ}^+(2632)$ state with mass $m_{D_{sJ}^{+}(2632)}=2.632~\mathrm{GeV}$~\cite{Evdokimov:2004iy}. It is important to note, however, that so-far the Selex discovery has not been confirmed by any other searches~\cite{Aubert:2004ku}. In $D$ decays, the situation is similarly ambiguous. Although the vector $D^{'*}$ resonance was discovered by Delphi~\cite{Abreu:1998vk} with a mass of
$m_{D^{'*}}=2.637~\mathrm{GeV}$ and spin-parity $1^-$, its existence
was not confirmed by other searches~\cite{Rodriguez:1998ng,Abbiendi:2001qp}. On the other hand recent theoretical studies~\cite{DiPierro:2001uu,Vijande:2003uk} indicate that both
radially excited vector states of $D$ as well as $D_s$ should have slightly
larger masses of $m_{D^{'*}}\simeq2.7~\mathrm{GeV}$ and $m_{D_{s}^{'*}}\simeq2.8~\mathrm{GeV}$~\cite{DiPierro:2001uu}. We use these theoretically predicted values in our analysis as well as for the radially excited pseudoscalar states, for which currently no experimental indications exist.
\par
In our  calculations we use for the heavy meson weak current coupling $\alpha = f_H \sqrt{m_H}$~\cite{Wise:1992hn,Becirevic:2002sc}, which we derive from the lattice QCD value of $f_{D} = 0.225~\mathrm{GeV}$~\cite{Wingate:2004xa} and experimental $D$ meson mass $m_{D}=1.87~\mathrm{GeV}$~\cite{Eidelman:2004wy} yielding $\alpha=0.31~\mathrm{GeV^{3/2}}$.  
The $\lambda$ coupling was usually~\cite{Casalbuoni:1996pg, Cheng:2004ru} determined from the value of $V(0)$. However, this derivation employed a single pole ansatz for the shape of $V(q^2)$. One can instead use data on $D^* \to D \gamma$ radiative decays. 
Following discussion in Refs.~\cite{Prelovsek:2000rj,Fajfer:1997bh},  using the  most recent data on $D^*$ radiative and strong decays~\cite{Eidelman:2004wy}, and accounting for the $SU(3)$ flavor symmetry breaking effects, we calculate $\lambda =- 0.526$ Gev$^{-1}$. 
The coupling $\beta \simeq 0.9$ has been estimated in Ref.~\cite{Isola:2003fh} relying on the assumption that the electromagnetic interactions of the light quark within heavy meson are dominated by the exchange of $\rho^0$, $\omega$, $\phi$ vector mesons.
\par
We fix the other free parameters ($\alpha_1, \alpha_2, \alpha', \tilde\alpha, \zeta, \mu, \tilde\zeta, \tilde\mu$) appearing in the HM$\chi$T Lagrangian and weak currents by comparing our model predictions to known experimental values of branching ratios $\mathcal B (D^0\rightarrow K^{*-}\ell^+\nu)$, $\mathcal B (D_s \rightarrow \phi\ell^+\nu)$, $\mathcal B (D^+\rightarrow \rho^0\ell^+\nu)$, $\mathcal B (D^+\rightarrow K^{0*}\ell^+\nu)$, as well as partial decay width ratios $\Gamma_L/\Gamma_T (D^+\rightarrow K^{0*}\ell^+\nu)$ and $\Gamma_+/\Gamma_- (D^+\rightarrow K^{0*}\ell^+\nu)$~\cite{Eidelman:2004wy}. In order to compare the results of our approach with experimental values, we calculate the decay rates for polarized final light vector mesons. Using helicity amplitudes $H_{+,-,0}$ defined in Sec. III and by neglecting the lepton masses we get~\cite{Bajc:1995km}:
\begin{equation}
\Gamma_a = \frac{G_F^2 m_H^2 |K_{HV}|^2}{96 \pi^3} \int_0^{y_m^V} y \mathrm{d}y  | H_a( y) |^2 |\vec p_V (y)|,
\end{equation} 
where $a=+,-,0$ and
\begin{equation}
y_m^V = \left(1-\frac{m_V}{m_H} \right)^2.
\end{equation} 
The constants $K_{HV}$ parametrize the flavor mixing relevant to a
particular transition, and are given in Table~\ref{PV_mixing_table}
together with the pole mesons.
\begin{table}
\caption{\label{PV_mixing_table} The pole mesons and the flavor mixing constants $K_{HV}$ for the $D\rightarrow V$ semileptonic decays.}
\begin{ruledtabular}
\begin{tabular}{cccccc}
$H$ & $V$ & $H^*$ & $H_P$ & $H_A$ & $K_{HV}$\\\hline
$D^0$ & $K^{*-}$ & $D^{*+}_s$, $D^{'*+}_{s}$ & $D_s^+$, $D_s^{'+}$ & $D_{sJ}(2463)^+$ & $V_{cs}$\\
$D^+$ & $\bar K^{*0}$ & $D^{*+}_s$, $D^{'*+}_{s}$ & $D_s^+$, $D_s^{'+}$ & $D_{sJ}(2463)^+$ & $V_{cs}$\\
$D^+_s$ & $\phi$ & $D^{*+}_s$, $D^{'*+}_{s}$ & $D_s^+$, $D_s^{'+}$ & $D_{sJ}(2463)^+$ & $V_{cs}$\\
$D^0$ & $\rho^-$ & $D^{*+}$, $D^{'*+}$ & $D^+$, $D^{'+}$ & $D_1(2420)$ & $V_{cd}$\\
$D^+$ & $\rho^0$ & $D^{*+}$, $D^{'*+}$  & $D^+$, $D^{'+}$ & $D_1(2420)$ & $-\frac{1}{\sqrt 2} V_{cd}$\\
$D^+$ & $\omega$ & $D^{*+}$, $D^{'*+}$  & $D^+$, $D^{'+}$ & $D_1(2420)$ & $\frac{1}{\sqrt 2} V_{cd}$\\
$D^+_s$ & $K^{*0}$ & $D^{*+}$, $D^{'*+}$  & $D^+$, $D^{'+}$ & $D_1(2420)$ & $V_{cd}$\\
\end{tabular}
\end{ruledtabular}
\end{table}
The transverse, longitudinal and total decay rates are then given trivially by
\begin{eqnarray}
\Gamma_T &=& \Gamma_+ + \Gamma_-,\nonumber\\*
\Gamma_L &=& \Gamma_0,\nonumber\\*
\Gamma_{~} &=& \Gamma_T + \Gamma_L.
\end{eqnarray} 
Consequently, the $A_0$ form factor does not contribute to any decay rate in this approximation and we can not fix the parameters $\alpha_2$ and $\tilde\zeta$ solely from comparison with experiment. Although $A_0$ actually does contribute indirectly through the relation~(\ref{PV_form_factor_relations}) at $q^2=0$ as manifested by Eq.~(\ref{eq_H0_A0}), this constraint is not automatically satisfied by our model. On the other hand, we can still enforce it "by hand" after the extrapolation to $q^2=0$ to obtain some information on these parameters. Due to the specific combinations in which the parameters appear in Eqs.~(\ref{eq_ff_HMcT}) we are further restrained to determining only the products $\tilde\alpha\tilde\mu$, $\alpha'\zeta$ and $\alpha'\mu$ using this kind of analysis. Lastly, since the only relevant contribution of $\alpha_1$ is to the $A_1$ form factor, we cannot disentangle it from the influence of $\alpha'\zeta$. Yet again we can impose the large energy limit relation~(\ref{eq_VA_LEET}) to extract both values independently.
\par
We calculate the result for $\tilde\alpha\tilde\mu$, $\alpha'\zeta$, $\alpha'\mu$ and $\alpha_1$ by a weighted average of values obtained from all the measured decay rates and their ratios taking into account for the experimental uncertainties. Furthermore, the values of $\alpha_1$ and $\alpha'\zeta$ are extracted separately by minimizing the fit function $(V(0)\xi - A_1(0))^2/(V(0)\xi + A_1(0))^2$. Both minimizations are performed in parallel and the global minimum is sought on the hypercube of dimensions $[-1,1]^4$ in the hyperspace of the fitted parameters. At the end we obtain the following values of parameters:
\begin{eqnarray}
\tilde\alpha\tilde\mu &=& 0.090~\mathrm{GeV}^{1/2} \nonumber\\*
\alpha'\zeta &=& 0.038~\mathrm{GeV}^{3/2} \nonumber\\*
\alpha'\mu &=& -0.066~\mathrm{GeV}^{1/2} \nonumber\\*
\alpha_1 &=& -0.128~\mathrm{GeV}^{1/2}
\end{eqnarray}
These values qualitatively agree with the analysis done in Ref.~\cite{Casalbuoni:1996pg} using a combination of quark model predictions and single pole experimental fits for all the form factors.  
\par
\begin{figure}[H]
\scalebox{0.9}{\includegraphics{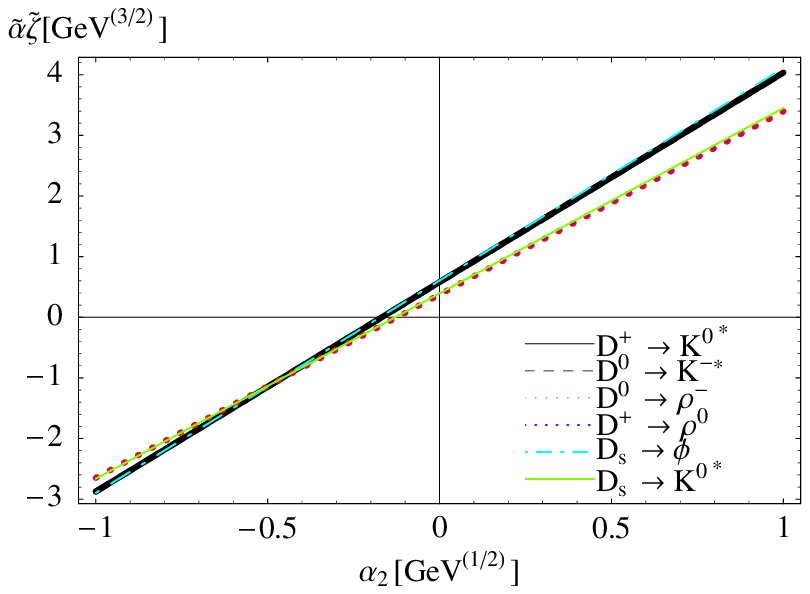}}
\caption{\label{fig_a2_tz} Solutions of Eq.~(\ref{PV_form_factor_relations}) in the $\alpha_2 \times \tilde\alpha\tilde\zeta$ parameter plane for the various decay channels considered.}
\end{figure}

We next use these values in relation~(\ref{PV_form_factor_relations}) to extract information on the the parameters $\alpha_2$ and $\tilde\zeta$. From Eqs.~(\ref{eq_ff_HMcT}) it is easy to see that the solutions lie on a straight line in the $\alpha_2 \times \tilde\alpha\tilde\zeta$ plane. We draw these for the various decay channels used in our analysis in Fig.~\ref{fig_a2_tz}.
We can see that all the decay channels considered fit approximately the same solution in the plane. Consequently, we can use any point on the approximate solution line to obtain the same prediction for the $q^2$ dependence of the $A_0$ form factor.
\par
It is important to note at this point that due to a high degree of interplay of the various HM$\chi$T parameters in the model predictions used in the fit, the values of the new model parameters obtained in such a way are very volatile to changes in the other inputs to the fit. Furthermore these are tree level parameter values and may in addition be very sensitive to chiral and $1/m_H$ corrections. Therefore their stated values should be taken {\it cum grano salis}. However more importantly, the form factor, branching ratio and polarization width ratio predictions based on this approach are more robust since they are insensitive to particular combinations of parameter values used, as long as they fit the experimental data. We estimate that chiral and heavy quark symmetry breaking corrections could still modify these predictions by as much as $30\%$.
\par
\begin{figure}[H]
\scalebox{0.9}{\includegraphics{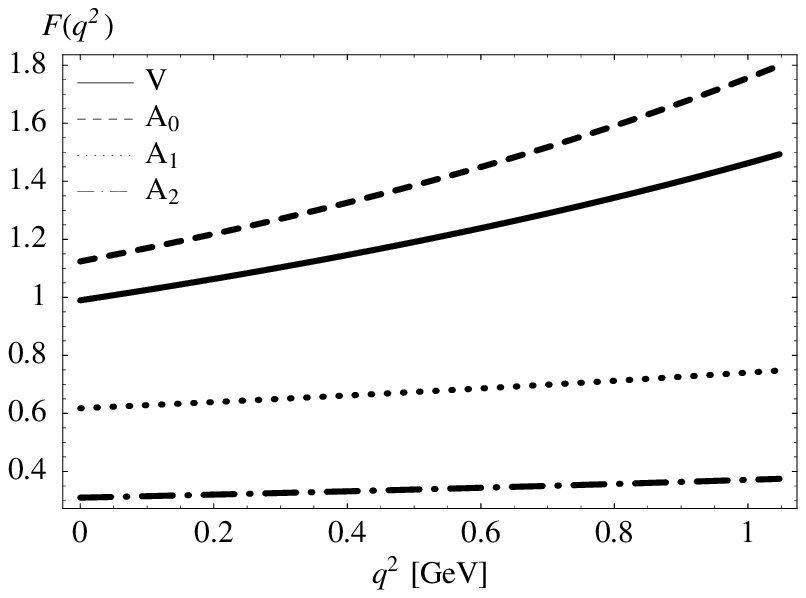}}
\caption{\label{fig_ff_d0k} Our model predictions for the $q^2$ dependence of the form factors $V(q^2)$ (solid line), $A_0(q^2)$ (dashed line), $A_1(q^2)$ (dotted line) and $A_2(q^2)$ (dash-dotted line) in $D^0\to K^{-*}$ transition.}
\end{figure}
\begin{figure}[H]
\scalebox{0.9}{\includegraphics{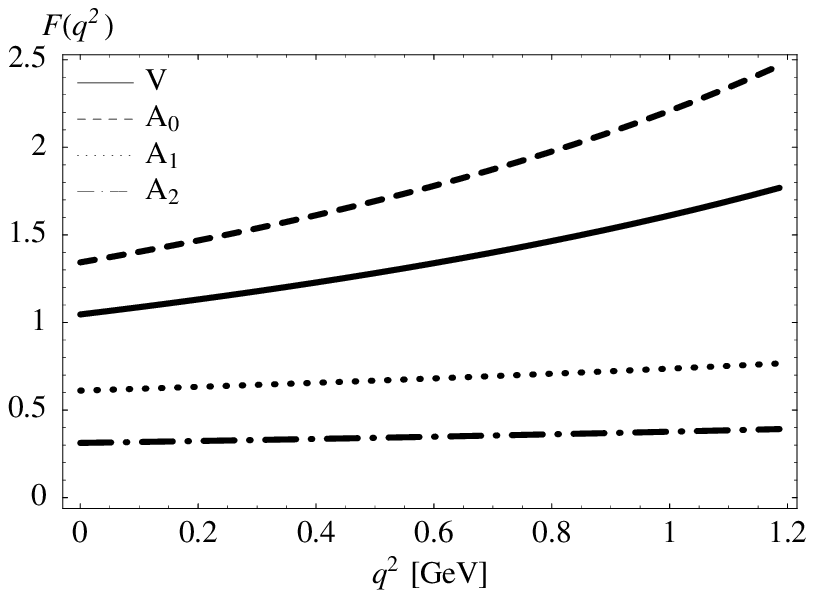}}
\caption{\label{fig_ff_d0rho} Our model predictions for the $q^2$ dependence of the form factors $V(q^2)$ (solid line), $A_0(q^2)$ (dashed line), $A_1(q^2)$ (dotted line) and $A_2(q^2)$ (dash-dotted line) in $D^0\to \rho^-$ transition.}
\end{figure}
\begin{figure}[H]
\scalebox{0.9}{\includegraphics{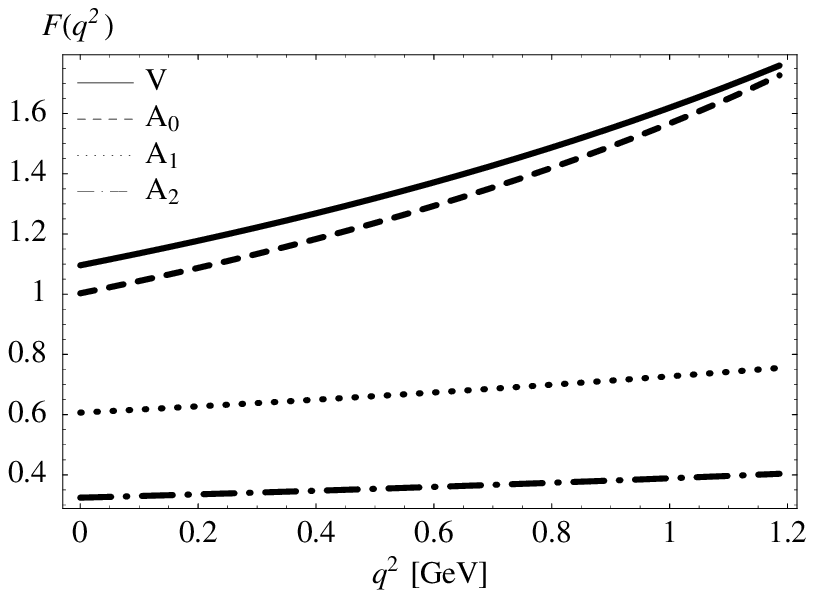}}
\caption{\label{fig_ff_dsphi} Our model predictions for the $q^2$ dependence of the form factors $V(q^2)$ (solid line), $A_0(q^2)$ (dashed line), $A_1(q^2)$ (dotted line) and $A_2(q^2)$ (dash-dotted line) in $D_s\to \phi$ transition.}
\end{figure}

We are now ready to draw the $q^2$ dependence of all the form factors for the $D^0\to K^{-*}$, $D^0\to \rho^-$ and $D_s\to \phi$ transitions. The results are depicted in Figs.~\ref{fig_ff_d0k},~\ref{fig_ff_d0rho}, and~\ref{fig_ff_dsphi}.
Our predictions for the shapes of the various form factors can also be summarized using the general formulas
\begin{eqnarray}
V(q^2) &=& \frac{V(0)}{(1-x)(1-ax)}, \nonumber\\
A_0(q^2) &=& \frac{A_0(0)}{(1-y)(1-a'y)}, \nonumber\\
A_1(q^2) &=& \frac{A_1(0)}{1-b'x}, \nonumber\\
A_2(q^2) &=& \frac{A_2(0)}{(1-b'x)(1-b''x)},
\label{eq_ff_general}
\end{eqnarray}
where as before $x=q^2/m_{H^*}^2$ and $y=q^2/m_{H_P}^2$. These expressions are actually simplifications of the form factor parametrizations~(\ref{eq_v_ff}), (\ref{eq_a0_ff}), (\ref{eq_a1_ff}) and (\ref{eq_a2_ff}) respectively. The parameters $V(0)$, $A_0(0)$, $A_1(0)$, $A_2(0)$, $a$, $a'$, $b'$ and $b''$, which we fix by nearest resonance saturation approximation and HM$\chi$T calculation at $q^2_{\mathrm{max}}$, are listed in Table~\ref{table_params} for the various decay channels considered.
\begin{table*}
\caption{\label{table_params} Our model predictions for the parameter values appearing in the general form factor formulas~(\ref{eq_ff_general}) for the various decay channels considered ($b''=0$ for all decay modes as explained in the text).}
\begin{ruledtabular}
\begin{tabular}{l|cccc|ccc}
Decay & $V(0)$ & $A_0(0)$ & $A_1(0)$ & $A_2(0)$ & $a$ &  $a'$ & $b'$ \\\hline 		$D^0\to K^{-*}$ & $0.99$ & $1.12$ & $0.62$ & $0.31$ & $0.57$ & $0.53$ & $0.74$  \\
	$D^0\to \rho^-$ & $1.05$ & $1.32$ & $0.61$ & $0.31$ & $0.55$ & $0.52$ & $0.69$ \\
	$D^+\to K^{0*}$ & $0.99$ & $1.12$ & $0.62$ & $0.31$ & $0.57$ & $0.53$ & $0.74$ \\
	$D^+\to \rho^0$ & $1.05$ & $1.32$ & $0.61$ & $0.31$ & $0.55$ & $0.52$ & $0.69$ \\
	$D^+\to \omega$ & $1.05$ & $1.32$ & $0.61$ & $0.31$ & $0.55$ & $0.52$ & $0.69$ \\
	$D_s\to \phi$ & $1.10$ & $1.02$ & $0.61$ & $0.32$ & $0.57$ & $0.53$ & $0.74$ \\
	$D_s\to K^{0*}$ & $1.16$ & $1.19$ & $0.60$ & $0.33$ & $0.55$ & $0.52$ & $0.69$ 
\end{tabular}
\end{ruledtabular}
\end{table*}
\par
Finally, we calculate branching ratios and partial decay width ratios for all relevant $D \to V \ell \nu_{\ell}$ decays. They are listed in Table~\ref{table_results} together with known experimentally measured values where we have marked those used in the fit of our model parameters.  
\begin{table*}
\caption{\label{table_results} The branching ratios and partial decay width ratios for the $D\rightarrow V$ semileptonic decays. Our model predictions and experimental results as explained in the text.}
\begin{ruledtabular}
\begin{tabular}{l|cc|cc|cc}
Decay & $\mathcal{B}$ (model) [\%] & $\mathcal{B}$ (Exp.) [\%] & $\Gamma_L/\Gamma_T$ (model) &  $\Gamma_L/\Gamma_T$ (Exp.) &  $\Gamma_+/\Gamma_-$ (model) &  $\Gamma_+/\Gamma_-$ (Exp.) \\\hline 	$D^0\to K^{-*}$ & $2.2$ & $2.15 \pm 0.35$~\cite{Eidelman:2004wy}\footnote{Values used in the fit of our model parameters} & $1.14$ & & $0.22$ & \\
	$D^0\to \rho^-$ & $0.20$ & $0.194\pm 0.039 \pm 0.013$~\cite{Blusk:2005fq} & $1.10$ & & $0.13$ & \\
	$D^+\to K^{0*}$ & $5.6$ & $5.73\pm 0.35$~\cite{Eidelman:2004wy}$^a$ & $1.13$ & $1.13\pm0.08$~\cite{Eidelman:2004wy}$^a$ & $0.22$ & $0.22\pm0.06$~\cite{Eidelman:2004wy}$^a$ \\
	$D^+\to \rho^0$ & $0.25$ & $0.25\pm 0.08$~\cite{Eidelman:2004wy}$^a$ & $1.10$ & & $0.13$ & \\
	$D^+\to \omega$ & $0.25$ & $0.17\pm0.06\pm0.01$~\cite{Blusk:2005fq} & $1.10$ & & $0.13$ & \\
	$D_s\to \phi$ & $2.4$ & $2.0\pm 0.5$~\cite{Eidelman:2004wy}$^a$ & $1.08$ & & $0.21$ & \\
	$D_s\to K^{0*}$ & $0.22$ &  & $1.03$ & & $0.13$ & 
\end{tabular}
\end{ruledtabular}
\end{table*}

\section{Conclusion}

We have investigated the $q^2$ dependence of the heavy to light form factors present in 
the $D \to V \ell \nu_{\ell}$ decays. First we have devised a general parametrization of the $H\to V$ form factors and then used it to compute the $D\to V$ form factors $q^2$ dependence in the framework of HM$\chi$T. Although we restrict our discussion to $D$ decays, the implications of the general form factor parametrization for the semileptonic decays of $B$ mesons are obvious, while also our HM$\chi$T calculations can easily be applied to these decays once more experimental data becomes available on excited $B$ meson resonances. Furthermore, the extension of our parametrization to tensor current form factors is trivial using HQET form factor relations~\cite{Isgur:1990kf,Burdman:1992hg}, which were found to hold even beyond the leading order in heavy quark mass expansion~\cite{Grinstein:2002cz,Grinstein:2004vb}.
\par 
In our calculations we do not include chiral corrections and $1/m_H$ 
corrections which might be important in charm decays as we already discussed in~\cite{Fajfer:2004mv}. As shown in Ref.~\cite{Boyd:1994pa} some of these corrections can in fact be absorbed into the leading order parameters while a sizable number of additional parameters remains undetermined. Since our approach is based on a global fit to existing experimental data on $D\to V$ semileptonic decays it does not seem possible at present to disentangle the effects of these new parameters and fix their values. Eventually, once more experimental results become available, it will be possible to learn more about some combinations of these parameters. We note however that while the values of HM$\chi$T parameters which are used and obtained in the fit of our model predictions may indeed be affected by chiral and $1/m_H$ corrections, the predicted shapes of the form factors are very robust since they are fixed by the underlying general form factor parametrization proposed in the text and by the masses of the involved resonances. Only the overall size of the individual form factors is determined by HM$\chi$T calculation at $q^2_{\mathrm{max}}$ and could thus be sensitive to chiral and $1/m_H$ corrections as well to the input parameters of the experimental fit. 
\par
The presence of charm meson resonances in our Lagrangian affects the values of the
form factors at $q^2_{\mathrm{max}}$ and induces saturation of the second poles in the parametrizations of the $V(q^2)$ and $A_0(q^2)$ form factors by the next radial excitations of $D_{(s)}^*$ and $D_{(s)}$ mesons respectively. The single pole  $q^2$ behavior of the $A_1(q^2)$ form factor is explained by the presence of a single $1^+$ state relevant to each decay, while in $A_2(q^2)$ in addition to these states one might also account for their next radial excitations. However, due to the lack of data on their presence we assume their masses being much higher than the first $1^+$ states and we neglect their effects.
\par
We point out that the single pole parametrization of all the form factors used in previous experimental studies cannot correctly satisfy known HQET and large energy limit constraints. In LEET~\cite{Charles:1998dr} and SCET~\cite{Hill:2004rx} studies the ratio of $V(0)$ and $A_1(0)$ obtained for $B \to \rho \ell \nu_{\ell}$  within these approaches was compared 
with experimental results for $D \to K^* \ell \nu_{\ell}$ which were obtained assuming such single pole dependence. 
\par
In the current literature there exist a number of studies of the $q^2$ shape of the $D\to V$ form factors. In the lattice simulation of Ref.~\cite{Abada:2002ie} the scaling behavior of the form factors has been properly included, but the pole/dipole fits of the form factors they used provide no physical information on the $q^2$ dependence of the form factors beyond the leading poles. However, our results for the $D \to K^* \ell \nu_{\ell}$ form factors are in good agreement with theirs. 
Similarly, the fits done in the quark model calculation of~\cite{Melikhov:2000yu} to their particular parametrization do not differ a lot from our $q^2$ behavior of the form factors.  
On the other hand, the authors of~\cite{Ball:1991bs} have used QCD sum rules to derive the $q^2$ 
dependence of the form factors. 
%They obtained for  $D \to K^* \ell \nu_{\ell} $ form factors at $q^2 =0$ that $V(0) = 1.1 \pm 0.25$, $A_1(0)=0.59 \pm 0.15$, $A_2(0) = 0.60 \pm 0.15$. 
As presented on Fig.~\ref{fig_ff_d0rho} and in Table~\ref{table_params} our results are in good agreement with theirs on values of $V(0)$ and $A_1(0)$ while our results for the $A_2(0)$ are somewhat lower.
% than theirs but agrees with lattice result for $D \to K^* l |nu_l$.
In Ref.~\cite{Ball:1993tp} the values of the form factors appearing in $D \to \rho \ell \nu_{\ell}$ decay have been investigated in the same approach and they
%the it was found that $V(0) = 1.0 \pm 0.2$, $A_1(0)=0.5 \pm 0.2$, $A_2(0) = 0.4 \pm 0.1$ what 
agree well with results of our model calculation and extrapolation.  
\par
We hope that the ongoing experimental studies will help to shed more light on the shapes of the $D\to V$ form factors.

\begin{acknowledgments}
We are greatly indebted to Damir Be\'cirevi\'c who originally initiated this work and provided useful insight and advice throughout the process of this study.
S. F. thanks  Alexander von Humboldt foundation for financial support
and A. J. Buras for his warm hospitality during her stay at
the Physik Department, TU M\"unchen, where part of this work has been done.
This work is supported in part by the Ministry of Higher Education, Science and Technology of the Republic of Slovenia.
\end{acknowledgments}

\bibliography{article}

\end{document}